\documentclass[conference]{IEEEtran}
\usepackage{epsfig}
\usepackage{times}
\usepackage{float}
\usepackage{afterpage}
\usepackage{amsmath}
\usepackage{amstext}
\usepackage{amssymb,bm}
\usepackage{latexsym}
\usepackage{color}
\usepackage{array,algpseudocode}
\usepackage{makecell}
\usepackage{diagbox}
\usepackage{amsmath}
\usepackage{amsthm}
\usepackage{graphicx}
\usepackage[center]{caption}
\usepackage{pstricks}
\usepackage{subfigure}
\usepackage{booktabs}
\usepackage{enumerate}
\usepackage[ruled]{algorithm}
\usepackage{romannum}
\usepackage{derivative}
\usepackage{cite}

\usepackage{bbm}
\def\BibTeX{{\rm B\kern-.05em{\sc i\kern-.025em b}\kern-.08em
    T\kern-.1667em\lower.7ex\hbox{E}\kern-.125emX}}
\usepackage[normalem]{ulem}
\usepackage{url}
\newtheorem{theorem}{Theorem}
\newtheorem{lemma}{Lemma}
\newtheorem{definition}{Definition}
\newtheorem{remark}{Remark}
\newtheorem{example}{Example}

\newtheorem{proposition}{Proposition}

\newcommand{\comment}[1]{}
\newcommand{\floor}[1]{\lfloor #1 \rfloor}
\newcommand{\ceil}[1]{\lceil #1 \rceil}
\let\svthefootnote\thefootnote
\newcommand\blankfootnote[1]{%
  \let\thefootnote\relax\footnotetext{#1}%
  \let\thefootnote\svthefootnote%
}

\allowdisplaybreaks

\begin{document}
\title{On the Capacity of ``Beam-Pointing'' Channels with Block Memory and Feedback: The Binary Case}
\author{\IEEEauthorblockN{Siyao Li}
\IEEEauthorblockA{\textit{ Communications and Information Theory Group (CommIT) } \\
\textit{ Technische Universit\"{a}t  Berlin }\\
Berlin, Germany \\
siyao.li@tu-berlin.de}
\and
\IEEEauthorblockN{Giuseppe Caire}
\IEEEauthorblockA{\textit{Communications and Information Theory Group (CommIT)} \\
\textit{ Technische Universit\"{a}t Berlin }\\
Berlin, Germany  \\
caire@tu-berlin.de} }
\maketitle

\begin{abstract}
Millimeter-wave (mmWave) communication is one of the key enablers for 5G systems as it provides larger system bandwidth and the possibility of packing numerous antennas in a small form factor for highly directional communication.  In order to materialize the potentially very high beamforming gain, the transmitter and receiver beams need to be aligned.
Practically, the Angle-of-Departure (AoD) remains almost constant over numerous consecutive time slots, which presents a state-dependent channel with memory. In addition, the backscatter signal can be modeled as a (causal) generalized feedback. The capacity of such channels with memory is generally an open problem in information theory. Towards solving this difficult problem, we consider a “toy model”,  consisting of a binary state-dependent (BSD) channel with in-block memory (iBM)~\cite{Kramer-memory} and one unit-delayed feedback. The capacity of this model under the peak transmission cost constraint is characterized by an iterative closed-form expression. We propose a capacity-achieving scheme where the transmitted signal carries information and meanwhile uniformly and randomly probes the beams with the help of feedback.  
\end{abstract}

\section{Introduction}
\label{sec:intro}
As compared to current 4G networks, 5G delivers significantly higher peak data rates and connection density. This performance improvement is achieved using a number of advanced techniques including exploiting millimeter wave (mmWave) frequency bands, advanced signal coding techniques, network slicing, etc. A fundamental technology to 5G's enhanced capacity and throughput is beamforming, which is used with phased array antenna systems to focus the wireless signal in a chosen narrow directional beam. This results in an improved signal at the user equipment (UE), and also less interference between the signals of individual UE \cite{RappaportTheodoreS2017OoMW,Javidi-AL,Scalable-song,mmPosterior}.
  Due to the narrow beams, communication is possible only when the transmitter and receiver beams are properly aligned \cite{Qiao2016}. 
Beam alignment (BA) is such a process to identify the optimal transmit/receive beam pair which attains the maximum received signal strength. In particular, in mmWave communications, the receiver SNR is ``above threshold’’ (i.e., a  given target spectral efficiency is achieved with very low packet error probability) in aligned conditions, and ``below threshold’’ (i.e., much lower than the required level to decode packets successfully) in non-aligned conditions. In this sense, the receiver SNR has a binary behavior.
In general, communication between the base station (BS) and the UE is coordinated using a technique known as beam sweeping, along with channel state information (CSI), which carries no information. 
Moreover, the feedback is ``protocol based'', i.e., it requires an explicit transmission from the user to the BS. 
An alternative or complementary approach consists of exploiting ``joint communication and sensing'' (JCAS) (see e.g.  \cite{ISAC-6G,ISAC-Tradeoff,ISAC-MAC,ISAC-IT} and references therein), i.e, based on the backscatter signal obtained from  the UE to the BS, the channel state, which in this case coincides with the angle-of-departure (AoD) of the line-of-sight path from the UE to the BS, can be estimated at the BS using radar-style processing. 
In the JCAS approach, there is no clear distinction between the sensing phase (training only) and the communication phase (data only). Instead, communication can take place 
while the state (the AoD) is being estimated via the backscatter signal. 

In response to the increasing demand for higher cost efficiency in wireless communications, many researchers have been devoted to setting up the theory on performance limits in the cost limited systems \cite{VerduS2002Seit,Sboui}.
In practical communication systems, the peak cost is limited by hardware restrictions. However, imposing the peak power constraint on the channel input makes the capacity characterization in certain channels become extremely difficult \cite{Peak-average}. Some works addressed the capacity of Gaussian (or Gaussian-like) channels for peak or both peak and average cost constraints \cite{SMITH1971203,Shamai-peak-avg1995,Dytso-updown2017,Chaaban-2018}, while in general, the capacity of channels with peak cost constraint is still an open problem.

 We investigate the JCAS scenario from an information-theoretic viewpoint. 
We consider a state-dependent channel with memory under a peak input cost constraint. 
The capacity of channels with memory and (generalized) feedback is typically an open problem in information theory. Nevertheless, an ``implicit'' capacity formula was given for channels with {\em in-block memory} (iBM)~\cite{Kramer-memory}, i.e., the state remains constant for blocks of $L$ time slots, 
and changes independent and identically from block to block. Such capacity formula is very hard to evaluate since it requires the optimization over length-$L$ sequences of 
conditional input distributions.

\paragraph*{Contribution} To emphasize the binary behavior of the received SNR and consider practical hardware cost limitations, we propose a simplified binary beam-pointing (BBP) channel model with iBM where the channel state characterizes the active beam at each channel use and remains the same in the whole block length of $L$. 
There are $M$ potential beams that can be aligned while only one of them provides the required beamforming gain. This channel state information is assumed to be perfectly known at the receiver (CSIR) but unknown at the transmitter.  
With one-channel-use-delayed feedback, the transmitter aims to detect the right beam direction and transmit messages efficiently and reliably. 
 In this work, we focus on the capacity of the BBP channel model. A closed-form expression of the capacity under a peak cost constraint is derived for a general block length $L$. An optimal iterative transmission strategy is proposed with input cost parameters updated based on the peak cost constraint, which can be directly applied to the JCAS scenario, where the transmitter sends messages and detects the state simultaneously. 

\paragraph*{Organization} The reminder of this paper is organized as follows. Section \ref{sec:form} introduces the system model and some definitions. Section \ref{sec:main} presents the main contributions. Section \ref{sec:examples} illustrates some numerical examples. 
Section  \ref{sec:conlude} concludes this work.

\paragraph*{Notations} For an integer $n$, we let $[n] = \{ 1, \cdots, n\}$ and $[n_1: n_2] = \{ n_1, \cdots, n_2\}$ for some integers $n_1 < n_2$.  $\underline{X}$ denotes a vector and $\underline{X}^n = [\underline{X}_1, \cdots, \underline{X}_n ]$ denotes a sequence of vectors. 
Let $\beta_{k}^i$ denote the  binary sequence with  length $i$ and the first one element appearing at the $k$-index (i.e., $\beta_2^i = \{01 \underbrace{\star \cdots \star}_{i-2} \}$ where $\star$ can be either 0 or 1).  $\mathbbm{1}{\{ \cdot \}}$ denotes an indicator function.    
 $| \mathcal A | $ represents the cardinality of a set $\mathcal A$. 
 Let $H( \cdot )$ denote the binary entropy function. We take $\log$ with base 2 throughout the paper. 

\section{System Model}
\label{sec:form}
We consider a BBP channel model with iBM~\cite{Kramer-memory}. The transmitter wishes to communicate a message to a receiver, over an iBM channel with a fixed state in each block,  and simultaneously estimate the state from one-unit-delayed feedback. 
For an iBM channel with block length $L$,  we consider $n$ total transmission time in channel uses where $n =  \ell L$ and $\ell$ is the number of blocks. Note that when $L=1$, the channel becomes the memoryless channel with independent states.
The channel state ${\underline S} \in \mathcal S$ is i) independent of the channel input, ii) remains constant for an interval of $L$ channel uses, and iii) i.i.d. according to $P_{\underline S}$ across the blocks. 
To  model a BS with $M$ antennas (possible transmission directions), we use ${\underline S} \in \{0, 1\}^M$ to denote the channel state with dimension $M$. In this work, ${\underline S}$ is a one-hot vector with the $m$-th element being one and zero otherwise. The index $m$ is a  random variable uniformly distributed over $[M]$ and is referred to as the {\it transmission direction},  i.e., the quantized AoD of the UE with respect to the BS array.

Let ${ \underline X}_{i,j} \in \mathcal X  := \{0,1\}^M$ denote the channel input, $Y_{i,j} \in \mathcal Y := \{0,1\} $ denote the channel output,  in block $i\in[\ell]$ at channel use $j\in[L]$. 
The received signal at channel use $j$ of  block $i$ is 
\begin{align}
Y_{i,j} = {\underline S}_i^T {\underline X}_{i,j}. \label{eq:channel-model}
\end{align} 
where ${\underline X}_{i,j}$ is a channel input vector  with dimension $M$, $Y_{i,j} $ is a binary scalar. Notice that $Y_{i,j} = 1$ if the single ``1’’ in $S_i$ coincides with a ``1’’ in $X_{i,j}$ and zero otherwise. This means that in order to maximize the output entropy, one must send equiprobable 0 and 1 in the transmission direction corresponding to the channel state.
The transmitter receives one-unit-delayed noiseless feedback from the receiver, i.e., $Y_{i,j}$ at the end of channel use $j$ of block $i$.
The joint probability distribution of the considered model is 
\begin{align}
&P_{W {\underline X}^n {\underline S}^{\ell} Y^n }( {\underline x}^n, {\underline s}^{\ell}, y^n, ) 
=P_W(w) \notag
\\&\times   \prod_{i=1}^{\ell} \left( P_{\underline S}( {\underline s}_i ) \prod_{j=1}^L P_{Y|{\underline X \underline S} }(  y_{i,j}  | {\underline x}_{i,j}  {\underline s}_i ) P({\underline x}_{i,j} | w, { y}_{i}^{j-1} ) \right) 
\label{eq:joint-distribution}
\end{align} 
where 
we denote $ y_{i}^{j-1}  = [y_{i,1}, \cdots, y_{i, j-1}]$.

\begin{definition}
A $(2^{ n R}, n, L)$ code for the BBP P2P channel with iBM, block length $L$ and  $\ell$ blocks under peak input cost constraint $B_\text{peak}$ consists of
\begin{itemize}
\item a discrete message set $\mathcal{W} = [1: 2^{nR} ]$;
\item a sequence of encoding functions $\phi_{i, j}: \mathcal{W} \times \mathcal{Y}^{iL+j-1} \to \mathcal{X}, i\in[\ell], j \in [L]$, such that $X_{i,j} = \phi_{i,j}(W, Y^{iL+j-1})$ subject to the following per-symbol peak cost constraint \begin{align}
b(X_{i,j}) \leq B_\text{peak} \label{eq:peak-input-cost}
 \end{align}
 where   $b: \mathcal X \to \mathbb{R}_+$ is an input cost function;
\item a decoding function $g: \mathcal{S}^{\ell} \times \mathcal{Y}^{n} \to \mathcal{W}$, such that $\hat{W} = g(Y^n, {S}^{\ell})$. The rate of such code is $R$ and the error probability is:
\begin{align}
P_e^{(n)} :=\frac{1}{2^{nR} } \sum_{w=1}^{ \lceil 2^{nR} \rceil} \mathbbm{P}( g(Y^n, {S}^{\ell}) \neq w | W = w) \label{eq:error-prob}
\end{align}
\end{itemize}
\end{definition}

\begin{definition}
A rate-peak-cost tuple $(R,  B_\text{peak})$ is said achievable if there exists a sequence (in $n$) of $(2^{nR}, n, L)$ codes that simultaneously satisfy 
\begin{subequations}
\begin{align}
\lim_{n\to \infty} P_e^{(n)} =0, 
\\
 b( {\underline X}_{i,j}) \leq B_\text{peak},\  \forall i \in[\ell], j\in[L]. \label{def:peak-cost}
\end{align}
\label{eq:Lcode-upper}
\end{subequations}
 where $b: \mathcal X \to \mathbb R_+$ is an input cost function.  
\end{definition}
The capacity of the BBP P2P channel with iBM under peak cost constraint $B_\text{peak}$ is the set of all achievable rates satisfying \eqref{eq:Lcode-upper}. 

 In this work, we consider $b(\cdot)$ to be the Hamming weight function (number of ones). 
 This is physically motivated by the fact that assuming constant transmission power per direction, the total transmission power is proportional to the number of directions in which $X_{i,j}$ sends a ``1’’.


\section{Main Results}
\label{sec:main}
The capacity of the channel model in \eqref{eq:channel-model}  is based on the multi-letter expression in the following Theorem. 
\label{subsec:converse}
\begin{theorem}
The capacity $C(B_\text{peak})$ of the BBP P2P channel with iBM, feedback and CSIR under peak cost constraint $B_\text{peak}$ is
\begin{align}
C(B_\text{peak}) &= \frac{1}{L} \max_{P_{{\underline X}^L|| Y^L} \in \mathcal P_{B_\text{peak}} }  H(Y^L | {\underline S} )
\end{align}
where  \begin{align} 
  P_{ {\underline X}^L \| Y^{L-1} }:= \prod_{i=1}^L P_{ { \underline X}_i | {\underline X}^{i-1}, Y^{i-1}}\label{eq:input-complicated}
   \end{align} and channel input satisfies 
the peak cost constraint in \eqref{def:peak-cost}. 
\label{thm-converse}
\end{theorem}

\begin{IEEEproof}
Achievability follows by random coding with optimized input distribution $P_{X^L}$. The converse follows directly from \cite[Theorem 1]{Kramer-memory}, and $H(Y^L | \underline X^L, \underline S ) =0$ as the output is a deterministic function of the channel state $\underline S$ and input $\underline X^L$.
\end{IEEEproof}

In this section, we derive a closed-form solution to the capacity for the BBP P2P channel with iBM under peak cost constraint $B_\text{peak}$. The optimal transmission strategy follows a JCAS  approach that implicitly estimates the channel state at the BS side. However, it is important and instructive to notice that the optimal strategy does not require that the state is perfectly known at the BS (as for example a beam sweeping protocol would do). The state is ``trapped’’ in a sufficiently small set of directions, such that transmitting over such directions maximizes the rate without wasting more time slots in refining the (unnecessary) knowledge of the state. 

\subsection{Converse}

For the capacity expressed in Theorem \ref{thm-converse}, the optimal input distribution satisfies the following property. 
 \begin{lemma} \label{lemma-bell}
Given some optimal input distribution
    up to time $t-1$, the input distribution beyond time $t-1$, 
    is optimal if and only if it maximizes
        \begin{align}
        \sum_{i=t}^L H(Y_i  | {\underline S}, y^{t-1}, Y_t^{i-1}). 
       \end{align}
\end{lemma}

\begin{IEEEproof}
         Express $H(Y^L| {\underline S} )$ as
   \begin{align}
&H(Y^L| {\underline S})  \notag
\\& =  \frac{1}{L} \sum_{i=1}^L H(Y_i  | { \underline S}, Y^{i-1}) \label{eq:newstage}
 \\& =  \frac{1}{L} \left( \sum_{i=1}^{t-1} H(Y_i  | {\underline S}, Y^{i-1}) + \sum_{i=t}^L H(Y_i  | {\underline S}, Y^{i-1})\right) \label{eq:sumH}
  \end{align}
We note that the distribution of the symbols $Y^{t-1}$ is not a function of the input distribution beyond time $t-1$ and $H(Y_i  | {\underline S}, Y^{i-1}) $ is a function of $P_{Y^{i-1}, {\underline S} } (y^{i-1}, {\underline s} )$ and $P_{Y_i | {\underline S}, Y^{i-1}} (y_i | {\underline s}, y^{i-1} )$. 
 Therefore, given the optimal input distribution up to time $t-1$, the input distribution beyond $t-1$ is optimal if and only if the sum $\sum_{i=t}^L H(Y_i  | {\underline S}, y^{t-1}, Y_t^{i-1})$ is maximized.  
\end{IEEEproof}

   By Bellman's principle of optimality \cite{DP}, \eqref{eq:sumH} is maximized when each of $H(Y_j | Y^{j-1}, \underline S)$ is maximized.
  The key technique used to derive the capacity upper bound is that the uniform distribution is entropy maximizing among all discrete distributions. Since the beam index is uniformly distributed over $[M]$, $H(Y_j | Y^{j-1}, \underline S)$ is maximized when $P_{Y^{j-1}, \underline S}( y^{j-1}, \underline s) = P_{Y^{j-1}, \underline S}( y^{j-1}, \underline s^{\prime}), \forall \underline s \neq \underline s^{\prime}$. 
   Recall that $\beta_i^j$ denotes the set containing all possible $j$-length binary sequences with the first non-zero element appearing at index $i$. From channel use $k$ on, the maximum rate $H(Y_j |y^{j-1} \in \beta_k^{j-1}, \underline S) =1, k < j \leq L$ is achieved when $P_{Y_j | Y^{j-1}, \underline S}(1 | y^{j-1} \in \beta_k^{j-1}, \underline S) = \frac{1}{2}$, as once a $Y_k =1$ is sent back, the transmission direction is detected. 
 Let  
 \begin{align}
c_k:= M \sum_{y^j \in \beta_k^j} P_{Y^j, \underline S}(y^j, \underline S).
   \end{align}
    \begin{subequations}
      \label{eq:peak-prob-notations}
   Then,  \begin{align}
     P_{Y^j, \underline S}(0^j, \underline S)  = 1 - \sum_{k=1}^{j }  \sum_{y^j \in \beta_k^j} P_{Y^j, \underline S}(y^j, \underline S)  =1 - \sum_{k=1}^{j } \frac{ c_k}{M},
     \\
    P_{Y_{j+1} | Y^j, \underline S}(1 | 0^j, \underline S)  =   \frac{ P_{Y^{j+1}, \underline S}( 0^j 1, \underline S) }{ P_{Y^j, \underline S}(0^j, \underline S) }
 = \frac{ c_{j+1} }{M - \sum_{k=1}^{j } c_k }.
     \end{align}
Plugging \eqref{eq:peak-prob-notations} into \eqref{eq:sumH}, we can express the capacity under peak cost constraint $B_\text{peak}$ in the following theorem.
     \end{subequations}
\begin{theorem} \label{th-peak-capacity-upper}
The capacity of the BBP channel with iBM and blocklength $L$ in \eqref{eq:channel-model} given one-unit-delayed feedback and CSIR under peak cost constraint  $B_\text{peak}$  is 
  \begin{align} 
 & C( B_\text{peak}) = \frac{1}{L} 
\sum_{j=1}^L  \left( 
  (1 -\frac{ \sum_{k=1}^{j-1} c_{k}}{M}  ) H( \frac{ c_{j} }{M - \sum_{k=1}^{j -1} c_k }) \right. \notag
   \\& \left. \qquad \qquad \qquad 
 +  \frac{\sum_{k=1}^{j-1} c_{k}}{M} \right), \label{eq:peak-each-rate}
  \end{align} 
  where $\sum_{k=1}^{0} c_{k} = 0$, $c_1 = \min( \frac{M}{2}, B_\text{peak})$, and
   \begin{align}
    & c_j = \min( \frac{M- \sum_{k=1}^{j-1} c_k }{2}, B_\text{peak} ), 1<j \leq L. \label{eq:b0s-peak}
  \end{align}  
\end{theorem}
\begin{IEEEproof} 
We prove the converse of this theorem by induction. The achievability of this theorem is shown later in Section \ref{subsec:achieve}. We first show that when $L=1$ the channel is equivalent to the channel without feedback, and the capacity is upper bounded by $H(Y_1 | \underline S) = H(\min( \frac{1}{2},  \frac{B_\text{peak} }{M} ) )$. 

  For an input vector, ${\underline x}$ with cost $c$, there are ${M \choose c}$ possible input candidates, which can be ordered in some predetermined manner. We use ${\underline c}_{l}$ to denote an input vector  ${\underline x}$ with cost $c$ and index $l$. 
  For the first input vector, let
\begin{align}
P_{{\underline X}_1}({\underline x}_1) = \begin{cases}
r_0 \ &{\underline  x_1} = \underline{0}
\\
r_{c} \alpha_{c, l} & { {\underline x}_1}  = {\underline c}_{l}, \forall l \in \left[ {M \choose c} \right], c \in[ B_\text{peak} ],
\end{cases} \label{eq:L=1-Bpeak}
\end{align}
where $\underline 0$ is a zero vector, $r_c = P_{b({\underline X}_1)}(b({\underline x}_1 ) = c)$ such that $\sum_{c \in [B_\text{peak}] } r_c =1$ and $\alpha_{c, l} = P_{ {\underline X}_1 | b({\underline X}_1) }(  {\underline c}_l |  c)$ such that $\sum_{ l \in \left[ {M \choose c} \right]} \alpha_{c, l} =1$ for all $c \in [B_\text{peak} ]$. 
As 
\begin{align}
P_{Y_1| {\underline S} }(y_1, \underline s) = \sum_{ {\underline x}_1} P_{ {\underline X}_1}( {\underline x}_1) \mathbbm{1}_{\{ y_1= {\underline s}^T {\underline x}_1 \} },\label{eq:L=1-Y|S}
\end{align}
combining  \eqref{eq:L=1-Y|S}  and \eqref{eq:L=1-Bpeak}, we have for any ${\underline s}$:
\begin{align}
P_{Y_1| {\underline S} }(y_1, \underline s) =
\begin{cases}
 (r_0 + \sum_{  {\underline x}_{1}  } r_c \alpha_{c, l} ) \mathbbm{1}_{ \{{\underline s}^T {\underline x}_{1} =0 \}}, &y_1 = 0 
\\ \sum_{  {\underline x}_{1} } r_c \alpha_{c, l}  \mathbbm{1}_{ \{{\underline s}^T {\underline x}_{1} =1\} }, &y_1 = 1.
\end{cases} \label{eq:L=1-Y|S-value}
\end{align}
Therefore,
\begin{align}
&  H(Y_1 | {\underline S})  
= \frac{1}{M} \sum_{ {\underline s} } H( \sum_{c=1}^{ B_\text{peak} }  \sum_{ l \in [ { M \choose c} ] }  r_c \alpha_{c, l} \mathbbm{1}_{ \{ {\underline s}^T {\underline x}_{1} =1\} } ). \label{eq:peak-L=1}
\end{align}
As CSI is not known at the transmitter and  the uniform distribution is entropy maximizing among
all discrete distributions, we have the term on the right-hand side of  \eqref{eq:peak-L=1} 
 is maximized when $ \sum_{c=1}^{ B_\text{peak} }  \sum_{ l \in [ { M \choose c} ] }  r_c \alpha_{c, l} \mathbbm{1}_{ \{ {\underline s}^T {\underline x}_{1} =1\} } \to \frac{1}{2}$  and  uniformly distributed for each ${\underline s}$. 
Considering the peak cost constraint, we obtain
\begin{align}
H(Y_1 | {\underline S})
& \leq H \left( \min(\frac{1}{2}, \frac{B_\text{peak} }{M} )  \right) \label{eq:L=1-expect-b-upper}
\end{align} 
where the equality in~\eqref{eq:L=1-expect-b-upper} holds by 
choosing $r_{ c} =1$ where $c = c_1$ in \eqref{eq:b0s-peak}, and randomly and uniformly the direction, i.e., $\alpha_{c, l} = \frac{1}{ {M \choose c } }$.  

Next,  we assume when $L=i$, $H(Y_i | Y^{i-1}, {\underline S})$ is upper bounded by \eqref{eq:peak-each-rate}, that is,
\begin{subequations}
\begin{align} 
&\sum_{y^{i-1} \in \beta_j^{i-1} } P_{Y^{i-1}, \underline S}(y^{i-1}, \underline S) =  \frac{ c_j }{M}, 
 \label{eq:peak-next-non-0}
 \\ &P_{Y^{i-1}, {\underline S}}(0^{i-1}, {\underline S}) = 1 - \frac{ \sum_{j=1}^{i-1} c_j }{M}, \label{eq:peak-next-0}
\\ &P_{Y_{i}| Y^{i-1}, {\underline S}}(0| y^{i-1} \in \beta_j^{i-1}, {\underline S}) = P_{Y_{i}| Y^{i-1}, {\underline S}}(1 | y^{i-1} \in \beta_j^{i-1}, {\underline S}) \notag
\\& \qquad \qquad \qquad \qquad  \qquad \qquad= \frac{1}{2}, \label{eq:Y-beta-half}
\\&P_{Y_{i}| Y^{i-1}, {\underline S}}(1 | 0^{i-1}, {\underline S}) = \min(\frac{1}{2}, \frac{ B_\text{peak} }{ M - \sum_{j=1}^{i-1} c_j } ),
\end{align} 
where $\{ c_j, \forall j \in[i] \}$ are given in \eqref{eq:b0s-peak}. 
\end{subequations}
 By assumption and Lemma \ref{lemma-bell}, we would like to show that when $L = i+1$,
\begin{align}  
H(Y_{i+1}| Y^i, {\underline S}) &\leq (1 -\frac{ \sum_{j=1}^{i} c_{j} }{M}  ) H( \min(\frac{1}{2}, \frac{ B_\text{peak} }{ M - \sum_{j=1}^{i} c_j}))  \notag
 \\& \qquad 
 +  \frac{\sum_{j=1}^{i} c_{j} }{M}.
 \end{align}  
$P_{Y^{i}, {\underline S}}(0^{i}, {\underline S}) = 1 - \frac{ \sum_{j=1}^{i} c_j }{M} $ can be verified by assumption, since
\begin{align} 
P_{Y^{i}, {\underline S}}(0^{i}, {\underline S}) &= P_{Y_{i} | {\underline S}, Y^{i-1}} (0 | {\underline S}, 0^{i-1} ) P_{ Y^{i-1}, {\underline S} }(0^{i-1}, {\underline S} ) \notag
\\& = \Big(1- \min(\frac{1}{2}, \frac{ B_\text{peak} }{ M - \sum_{j=1}^{i-1} c_j }) \Big)\!(1- \frac{ \sum_{j=1}^{i-1} c_j }{M})  \notag
\\& = 1 - \frac{ \sum_{j=1}^i c_j}{M},  \label{eq:Y^i=0^i}
\end{align} where $c_i =\min( \frac{M- \sum_{j=1}^{i-1} c_j}{2}, B_\text{peak} )$. 
As 
\begin{align*} 
P_{Y^{i}, {\underline S}}(0^{i}, {\underline S}) + \sum_{k=1}^{i} \sum_{y^i \in \beta_k^i} P_{Y^{i}, {\underline S}}(y^{i}, {\underline S})=1,
\end{align*} 
 we obtain 
 \begin{align*} 
 \sum_{y^{i} \neq 0^{i} } P_{Y^{i}, {\underline S}}(y^{i}, {\underline S}) =  \frac{ \sum_{j=1}^{i} c_j }{M}.
\end{align*} 
Furthermore,
  \begin{align}
  & H(Y_{i+1} | Y^{i}, {\underline S} ) \notag
   \\&=   \sum_{ {\underline s} } \sum_{ y^i}  \left(  P_{Y^{i}, S}(0^{i}, {\underline s} ) H(Y_{i+1} | Y^i= 0^{i}, {\underline s} ) \right. \notag
  \\& \left. + \sum_{j=1}^i P_{Y^{i}, {\underline S} }( \beta_{j}, {\underline s} ) H(Y_{i+1} |  Y^i=\beta_j^i, {\underline s} )   \right)  \notag
   \\& \leq  \sum_{ {\underline s} } \sum_{ y^i} P_{Y^{i}, {\underline S} }( 0^{i}, {\underline s} )  H(Y_{i+1} | Y^i= 0^{i}, {\underline s} )\label{eq:H-Y_L+1-p1}
 %
 %
  \\&  +  \frac{\sum_{j=1}^i c_j}{M}. \label{eq:entropy-Y_L+1}
   \end{align}    
 \eqref{eq:entropy-Y_L+1} holds since $P_{Y^{i}, S}(  y^{i}, {\underline s} ), \forall  y^{i} \in \{0,1\}^{i} $ do not depend on the optimal input distribution at channel use $i+1$ and have been determined by the optimal input distribution by channel use $i$, and $H(Y_{i+1} |  \beta_j^i, {\underline s}) =1$  follows by \eqref{eq:Y-beta-half}. 
 Consider the terms in \eqref{eq:H-Y_L+1-p1}. $ H(Y_{i+1} | Y^i=0^i, {\underline S} )$ is maximized when $P_{Y_{i+1}| Y^i, {\underline S}}( Y_{i+1}=1 | Y^i = 0^i, {\underline s} )$ is uniformly distributed for each ${\underline s}$. It is also monotonically increasing when $P_{Y_{i+1}| Y^i, {\underline S}}( Y_{i+1}=1 | Y^i = 0^i, {\underline S} ) \in [0, \frac{1}{2}]$.    According to the peak cost constraint, we have
  \begin{align*}  
  P_{Y_{i+1}, Y^i, {\underline S}}( Y_{i+1}=1, Y^i = 0^i, {\underline S} ) &\leq P_{Y_{i+1}, {\underline S}}( Y_{i+1}=1, {\underline S} ) \notag
  \\ & \leq \frac{ B_\text{peak}}{M},
      \end{align*}  and
       \begin{align}  
  &P_{Y_{i+1} |Y^i, {\underline S}}( Y_{i+1}=1| Y^i = 0^i, {\underline S} ) \notag
  \\&=\frac{ P_{Y_{i+1}, {\underline S}}( Y_{i+1}=1, Y^i = 0^i,{\underline S} ) }{P_{Y^{i}, {\underline S}}(0^{i}, {\underline S}) } \notag
  \\ & \leq  \frac{ \frac{ B_\text{peak}}{M}}{ 1 - \frac{ \sum_{j=1}^i c_j }{M}} \notag
  \\& = \frac{ B_\text{peak} }{ M - \sum_{j=1}^{i} c_j }. \label{eq:last-x}
      \end{align} 
   Plugging \eqref{eq:Y^i=0^i} and \eqref{eq:last-x} into \eqref{eq:H-Y_L+1-p1} and \eqref{eq:entropy-Y_L+1}, we have
        \begin{align}
  & H(Y_{i+1} | Y^{i}, {\underline S} ) \notag
   \\& \leq (1 -\frac{\sum_{j=1}^{i} c_j }{M}  ) H( \min(\frac{1}{2}, \frac{ B_\text{peak} }{ M - \sum_{j=1}^i c_j})) +  \frac{\sum_{j=1}^i c_j }{M}. \label{eq:max-L+1}
   \end{align} 
   Therefore, we conclude the capacity is upper bounded by \eqref{eq:peak-each-rate}.
\end{IEEEproof} 
\subsection{Achievability}
\label{subsec:achieve}

In this subsection, we prove that Theorem \ref{th-peak-capacity-upper} can be achieved by applying the transmission strategy illustrated in Algorithm \ref{alg:JSAC-peak} in each block.

 Algorithm \ref{alg:JSAC-peak} is inspired by the iterative expression of $c_i$ in \eqref{eq:b0s-peak}. As mentioned before, to maximize the achievable rate, we must send equiprobable 0 and 1 in the transmission direction. With the help of feedback, although the transmission direction is not known at the transmitter, we can recursively trap it in a smaller set to satisfy the limited peak cost constraint. Let ${\mathcal B}_{y^i}$ denote the set of beam indices containing the transmission direction at channel use $i$ when channel output $Y^i = y^i$. Let $\mathcal B_i^e$ denote the set of beam indices to be explored at channel use $i$.   We initialize ${\mathcal B}_{y^0} = [M]$, ${\mathcal B}_{0}^e = \emptyset$, and a sequence of $\{ c_1, \cdots, c_L\}$ iteratively solved by \eqref{eq:b0s-peak}. At the beginning of channel use $i$, we update ${\mathcal B}_{y^i}$ and choose some number of beam indices randomly and uniformly from ${\mathcal B}_{y^{i}}$ based on the casual feedback $Y_{i-1}$. Specifically,  we use $k, k\in[L]$ to record the number of channel uses until the transmitter selected the ``right" directions (i.e., $Y_{k}=1$). Before that, the transmitter randomly and uniformly chooses $c_i, i\leq k$ beam indices from ${\mathcal B}_{y^{i-1}}$. After that,  the transmitter randomly and uniformly chooses $\frac{c_k}{2^{i-k}}, i>k$ beam indices from ${\mathcal B}_{y^{i-1}}$. These selected beam indices are stored in set $\mathcal B_i^e$. 
 
\begin{algorithm}[h] 
\caption{JCAS Scheme} \label{alg:JSAC-peak}
\begin{algorithmic}[1]
 \State {\bf Initialization:} 

1) Let $Y^0 = 0$, ${\mathcal B}_{y^0} = [M]$ and  $\mathcal B_0^e = \emptyset$.

2) Given a sequence of $\{ c_1, \cdots, c_L\}$ by \eqref{eq:b0s-peak}.

\State {\bf Recursions:}  
  \For{$ i = 1: L$}  
  	 \If {$Y_{i-1} = 0$}
	  \State ${\mathcal B}_{y^{i}} = {\mathcal B}_{y^{i-1}} \backslash \mathcal B_{i-1}^e$.
  \State Randomly and uniformly choose $c_i$ beam indices from ${\mathcal B}_{y^{i}}$, and the selected directions are stored in $\mathcal B_i^e$.	
	 \Else  
	  \State $k=i$.
	  \State ${\mathcal B}_{y^{i}} =\mathcal B_{i-1}^e$.
	   \State Randomly and uniformly choose $\frac{c_k}{2^{i-k}}$ beam indices from ${\mathcal B}_{y^{i}}$, and the selected directions are stored in $\mathcal B_i^e$.
	 	\EndIf
		
%
	%
		
\EndFor
\end{algorithmic}
\end{algorithm}


One can easily verify that when $L=1$ the rate
\begin{align*}  
R_1 = H(\min( \frac{1}{2},  \frac{B_\text{peak} }{M} ) )
\end{align*}
 is achievable by Algorithm \ref{alg:JSAC-peak}. 
When $L \geq 2$, at channel use $i \in[ 2, L)$, based on the feedback,
\begin{itemize}
\item if $Y_{i-1} = 0,$ it means that the transmission direction is not within the beam indices selected by $\underline X_{i-1}$. We simply exclude these selected beam indices from ${\mathcal B}_{y^{i-1}} $, and update ${\mathcal B}_{y^{i}} = {\mathcal B}_{y^{i-1}} \backslash \mathcal B_{i-1}^e$. 
Then, we apply the same strategy but with $| {\mathcal B}_{y^{i-1}}| - c_{i-1}$ possible directions to be explored.
\item if $Y_{i-1} = 1$, the right transmission direction is detected in some known set and the peak cost constraint is satisfied already. We can repeat partitioning this set into two halves and sending messages in the directions of each half with equal probability to achieve the maximum rate $R =1$.
\end{itemize}

Since the channel state is uniformly distributed, according to the transmission strategy in Algorithm \ref{alg:JSAC-peak}, we obtain \eqref{eq:peak-prob-notations}. 
    Hence, the achievable rate when $Y^{i-1} = 0^{i-1}$ is
      \begin{align*}
    &P_{Y^{i-1}, \underline S}(0^{i-1}, \underline S) H(P_{Y_i | Y^{i-1}\underline S}(1 | 0^{i-1}, \underline S)) \notag
   \\& \!\!\!\!\!\!\!\! = (1- \frac{ \sum_{k=1}^{i-1} c_k }{M}) H(\frac{ c_i }{ M- \sum_{k=1}^{i-1} c_k }).
  \end{align*}   
  Also, once the channel feeds back a $Y_k =1$, we obtain \eqref{eq:Y-beta-half} for $i >k$,
which gives
   \begin{align*}
   H(Y_i | y^{i-1} \in  \beta_k^{i-1},  \underline S ) =1,  \forall k \leq i-1.
  \end{align*}   
Therefore, at channel use $i\geq 2$, the information rate 
 \begin{align*}
 R_i &= P_{Y^{i-1}, \underline S}(0^{i-1}, \underline S) H( P_{Y_i | Y^{i-1}\underline S}(1 | 0^{i-1}, \underline S)) \notag
 \\&  \qquad + (1 - P_{Y^{i-1}, \underline S}(0^{i-1}, \underline S) ) H(\frac{1}{2})
 \\& = (1 -\frac{ \sum_{k=1}^{i-1} c_{k} }{M}  ) H( \frac{ c_i }{ M - \sum_{k=1}^{i-1} c_k }))   +  \frac{\sum_{k=1}^{i-1} c_{k} }{M}.  
\end{align*}
The overall average rate is
$ R = \frac{1}{L} \sum_{i=1}^L R_i$, which is \eqref{eq:peak-each-rate}. 

As introduced above, the transmitter identifies the channel state based on the feedback (sensing phase)  and simultaneously transmits information based on the sensing result (communication phase). There is no clear distinction between communication and sensing.
 Therefore, the transmission strategy in Algorithm \ref{alg:JSAC-peak} follows a JCAS manner.

Note that Algorithm \ref{alg:JSAC-peak} is one of the capacity-achieving transmission strategies. 
The optimal input sequences and distributions under peak cost constraint $B_\text{peak}$ are not unique and can be referred to in the following example.

\begin{example} $L=1, M = 6$ and $B_\text{peak} = 2$. 
\end{example}
The capacity is 
\begin{align*}
C_\text{NoEst}(B_\text{peak}=2) = H( \frac{B_\text{peak}}{M}) = H(\frac{1}{3}). 
\end{align*}
This can be achieved by taking 
\begin{align*}
P_{{\underline X}_1}( 110000) = P_{{\underline X}_1}( 001100) = P_{{\underline X}_1}( 000011) = \frac{1}{3},
\end{align*}
and can also be achieved by taking
\begin{align*}
&P_{{\underline X}_1}( 110000) = P_{{\underline X}_1}( 000110)  =  P_{{\underline X}_1}( 101000) 
\\ &= P_{{\underline X}_1}( 001100)  
= P_{{\underline X}_1}( 010001) = P_{{\underline X}_1}( 000011)  = \frac{1}{6}.
\end{align*}
As long as $\underline X_1$ satisfies the peak cost constraint and $\sum_{ {\underline x}_1 }P( {\underline s}^T {\underline x}_1 = 1)$ is uniformly distributed for all ${\underline s}$, we can achieve the capacity.

In Theorem \ref{th-peak-capacity-upper}, we assume the number of chosen directions as some integers, i.e., $M = 2^n$ and $B_\text{peak}\in \mathbb{Z}_+$. For some other values of $M$ or  $B_\text{peak}$ resulting $c_i$ in Theorem \ref{th-peak-capacity-upper} to be some non-integer values,  we simply partition the set into two parts with equal probability of having $\floor{c_{i} }$ and $\floor{c_{i} }+1$ disjoint elements if  $c_{i}$ is not an integer in Algorithm \ref{alg:JSAC-peak}. 

\section{Numerical Examples}
\label{sec:examples}
\begin{figure}[t]
    \centering 
    \includegraphics[width=1\columnwidth]{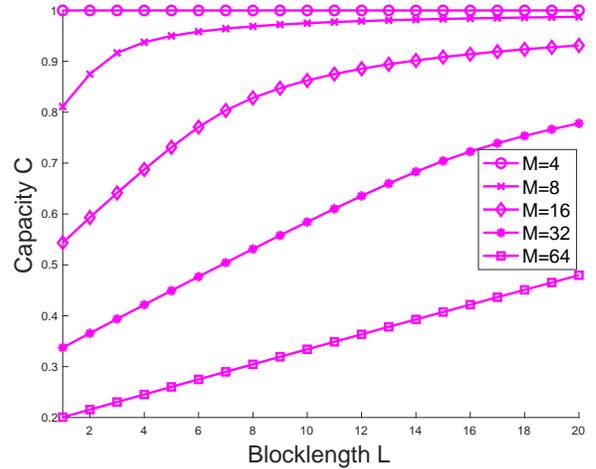}
    \caption{Capacity vs. blocklength under fixed $B_\text{peak}=2$.}
    \label{fig:BvsC} 
\end{figure}

In this section, we evaluate the capacity result in Theorem \ref{th-peak-capacity-upper} via some numerical examples. In particular, we investigate the relationship between capacity and blocklength under different peak input cost constraints in Fig. \ref{fig:LvsC} and different numbers of beams in Fig. \ref{fig:BvsC} separately.  

\subsection{Fixed number of directions}
\begin{figure}[t]
    \centering 
    \includegraphics[width=1\columnwidth]{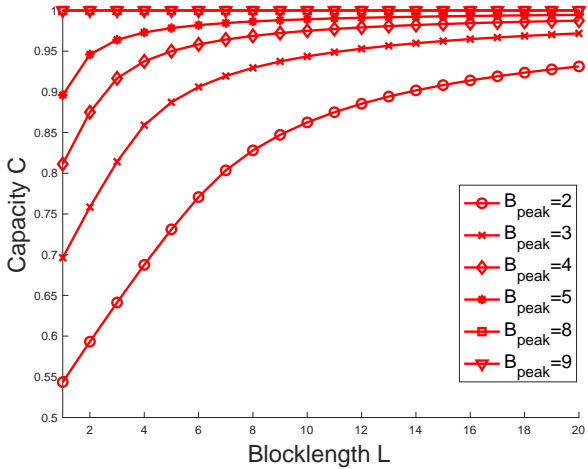}
    \caption{Capacity vs. blocklength under fixed $M=16$.}
    \label{fig:LvsC} 
\end{figure}

We fix the number of possible beams $M=16$ and vary the peak cost constraint $B_\text{peak} = \{2,3,4,5,8,9\}$ in Fig. \ref{fig:LvsC}. We skip $B_\text{peak} = \{6, 7\}$, but one can imagine the curve trend will be  similar to $B_\text{peak}=5$ and get closer to $B_\text{peak}=8$. It is intuitive that capacity $C$ is a monotonically nondecreasing function of the blocklength $L$ 
under the same cost $B_\text{peak}$, since the probability of successfully detecting the transmission directions rises as $L$ enlarges.  
In addition, capacity $C$ is a monotonically nondecreasing function of the peak cost constraint $B_\text{peak}$ under the same number of blocklength $L$, following a similar reason as just mentioned. We also observe that the capacity results ($C=1$) coincide when $B_\text{peak} \geq \frac{M}{2} = 8$, as the transmitter is able to ensure the received signal $Y=1$ and $Y=0$ with equal probability at each channel use. 
The capacity gap under different peak cost constraints is relatively large when the number of blocklength is small. While the gap becomes smaller and smaller as the number of blocklength increases since the transmission direction will eventually be detected and the transmitter can send coded messages with maximum achievable rate $R=1$. 

\subsection{Fixed peak cost constraint}
We fix the peak input cost constraint $B_\text{peak}=2$ and vary the number of beams $M=\{4, 8, 16, 32, 64\}$ in Fig. \ref{fig:BvsC}. 
The capacity $C=1$ is achieved when $M \leq 2B_\text{peak} =4$ for all blocklength $L$ as the transmitter is able to ensure the received signal $Y=1$ and $Y=0$ with equal probability at each channel use. The capacity $C$ is a monotonically nonincreasing function of the number of beams $M$ under the same number of blocklength $L$, since the transmitter is less likely to detect the channel state for a large number of beams when the peak input cost is limited. Together with Fig. \ref{fig:LvsC}, these two figures indicate the tight relationship between communication and sensing. That is, the more easily to detect the channel state, the more efficiently to transmit information.

\section{Conclusion}
\label{sec:conlude}
This work investigated the  binary beam-pointing channel with in-block memory and feedback that captures the main feature of the beam alignment problem in mmWave communications while being sufficiently simple to be analyzed in information theory. In addition, we take practical hardware restrictions (i.e., peak input cost constraint) into consideration. 
We characterized the capacity of this simplified channel model in a closed form and presented a general iterative scheme performing joint communication and sensing based on the peak cost constraint. Simulation results further demonstrate the relationship between blocklength and capacity under different sizes of beams and peak cost constraints. This study offers some useful perspectives on how to approach beam alignment issues from an information theory viewpoint, and it can be applied to a wider range of situations. However, this extension is typically highly non-trivial. 
Nevertheless, we believe that the guidelines for system design arising from the clean information-theoretic treatment of this model already clearly emerge and are somehow unexpected.  For example, our results clearly show that performing separately the functions of state estimation (e.g., beam sweeping) and communication, in distinct probing (estimation-only) and data (communication-only)  is suboptimal. Instead, the optimal strategy consists of a sort of ``estimating while communicating’’ joint scheme, fully embracing the principle of JCAS. Furthermore, the exact determination of the channel state at the transmitter is not generally a requirement of the optimal strategy in terms of communication, where the state is determined up to some uncertainty. 

\bibliographystyle{IEEEtran}
\bibliography{refs-isac}
\end{document}